\documentclass[aps,floatfix,twocolumn,showpacs]{revtex4}

\usepackage{natbib}
\usepackage{graphics,bm}
\usepackage{epsfig}
\usepackage{amsmath}
\usepackage{amssymb}
\usepackage{hyperref}
\newcommand{\be}{\begin{equation}}
\newcommand{\ee}{\end{equation}}
\newcommand{\bea}{\begin{eqnarray}}
\newcommand{\eea}{\end{eqnarray}}
\newcommand{\rr}{\mathbf{r}}
\newcommand{\kk}{\mathbf{k}}

\begin{document}

\title{Spin squeezing in Bose-Einstein condensates: Limits imposed by decoherence and non-zero temperature}

\author{Alice Sinatra}
\affiliation{Laboratoire Kastler Brossel, Ecole Normale Sup\'erieure, UPMC and CNRS,  Paris, France}

\author{Jean-Christophe Dornstetter}
\affiliation{Laboratoire Kastler Brossel, Ecole Normale Sup\'erieure, UPMC and CNRS,  Paris, France}

\author{Yvan Castin}
\affiliation{Laboratoire Kastler Brossel, Ecole Normale Sup\'erieure, UPMC and CNRS,  Paris, France}

\begin{abstract}
We consider dynamically generated spin squeezing in interacting bimodal condensates.
We show that particle losses and non-zero temperature effects in a multimode theory completely change the scaling of the best squeezing 
for large atom numbers. We present the new scalings and we give approximate analytical expressions for the squeezing
in the thermodynamic limit. Besides reviewing our recent theoretical results, we give here a simple physical picture of how decoherence
acts to limit the squeezing. We show in particular that under certain conditions the decoherence due to losses and non-zero temperature 
acts as a simple dephasing.
\end{abstract}                                                                 

\date{\today}
\maketitle
\tableofcontents

\section{Introduction}
Spin squeezing is about creating quantum correlations in a many-body system that can be useful for metrology. 
An example is that of atomic clocks, where the ultimate signal to noise ratio, once all the technical noise has been eliminated, 
can be improved by manipulating and controlling the system at the level of its quantum fluctuations.

\subsection{Spin squeezing and atomic clocks}
\begin{figure}
\centerline{\includegraphics[width=8cm,clip=]{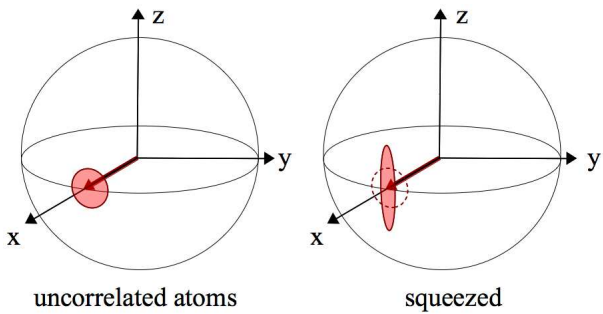}}
\caption{(Color online) Uncorrelated state and squeezed state represented on the Bloch sphere.
\label{fig:spin}}
\end{figure}

The aim of an atomic clock is to measure precisely the energy difference between two atomic states $a$ and $b$
that are for example two hyperfine states of an alkali atom. To explain how the clock works and to introduce 
spin squeezing, we shall describe the ensemble of $N$ atoms used in the clock
using the picture of a ``collective spin" that evolves on the so-called Bloch sphere.
The collective spin is simply the sum of the effective spins $1/2$ that describe the internal degrees of freedom of each atom.
In the second quantized formalism the three hermitian spin components $S_x$, $S_y$ and $S_z$ are defined by:
\be
S_x+i S_y={a^\dagger b} \,, \hspace{0.5cm} S_z={a^\dagger a - b^\dagger b \over 2}\,,
\label{eq:SxSySz}
\ee
where $a^\dagger$ and $b^\dagger$ are creation operators of particles in the internal states $a$ and $b$. 
The spin operators are dimensionless and obey the commutation relations $[S_x,S_y]=i S_z$ and cyclic permutations.
 For the moment we do not care about 
the external degrees of freedom of the atoms.
The component $S_z$ of the collective spin is half the population difference between states $a$ and $b$, while $S_x$ and $S_y$ 
describe the coherence between these states. If the $N$ atoms are prepared in a coherent superposition of states
$a$ and $b$ with relative phase $2 \phi$:
\be
|\phi\rangle_N = {1 \over \sqrt{N!}} \left( {e^{i\phi}a^\dagger  + e^{-i\phi}b^\dagger \over \sqrt{2}}\right)^N |0\rangle \,,
\label{eq:phase_state}
\ee
where $|0\rangle$ is the vacuum,
the collective spin lies on the equatorial plane of the Bloch sphere, pointing at an angle $-2\phi$ with respect to the $x$ axis. 
For non-interacting atoms, the further evolution is ruled by the Hamiltonian
\be
H_0=\hbar \omega_{ab} S_z \label{eq:H0}
\ee
and the spin precesses around the $z$ axis with the Larmor frequency $\omega_{ab}$. The atomic clock measures the phase accumulated 
by the collective spin during a long precession time $\tau$. From this phase the frequency $\omega_{ab}$ is deduced. Atomic
clocks are nowadays so precise that they are sensitive to the quantum noise of the collective spin. If for example the spin is initially prepared
along the $x$ axis (in an eigenstate of $S_x$), it has necessarily fluctuations in the transverse components $S_y$ and $S_z$ such that:
\be
\langle S_x \rangle={N\over 2} \,, \hspace{0.5cm} \Delta S_y \Delta S_z \ge {1 \over 2} |\langle S_x \rangle| \,.
\label{eq:Heisenberg}
\ee
In particular fluctuations of $S_y$ introduce a statistical variance on the accumulated phase during the 
precession time and on the measured frequency $\omega_{ab}$. In the uncorrelated state (\ref{eq:phase_state}) with $\phi=0$, 
root mean square fluctuations of $S_y$ and $S_z$ are equal: $\Delta S_y=\Delta S_z=\sqrt{N}/2$. In a Ramsey measurement with interrogation time 
$\tau$, these quantum fluctuations introduce the root 
mean square fluctuations of the measured frequency equal to \cite{Wineland:1994}:
\be
\Delta \omega_{ab}^{\rm unc} = {1 \over \sqrt{N} \tau} \,.
\ee
This noise coming from quantum fluctuations, intrinsic to the initial state where each atom is in a superposition of $a$ and $b$,
is known in clocks as the ``partition noise". The idea of spin squeezing \cite{Ueda:1993} is that
the Heisenberg relation (\ref{eq:Heisenberg}) allows to reduce $\Delta S_y$ 
provided that $\Delta S_z$ is increased. This idea is 
illustrated in Fig.\ref{fig:spin}. 
To quantify the spin squeezing we use the parameter $\xi^2$ introduced in  
\cite{Wineland:1994}:
\be
\xi^2={N \Delta S_{\perp,{\rm min}}^2 \over |\langle {\bf S} \rangle|^2} \label{eq:xi2}
\ee
where  $N$ is the total atom number, $\Delta S_{\perp,{\rm min}}^2$ is the minimal variance of the spin orthogonally to its mean value
$\langle {\bf S} \rangle$. The state is squeezed if and only if $\xi^2<1$. 
As explained in \cite{Wineland:1994}, $\xi$ directly gives the reduction of the statistical
fluctuations of the measured frequency $\omega_{ab}$ with respect to uncorrelated atoms, for the same atom number $N$ and 
the same Ramsey time $\tau$:
\be
\Delta \omega_{ab}^{\rm sq}= \xi \Delta \omega_{ab}^{\rm unc}={\xi \over \sqrt{N} \tau} \,.
\ee
The parameter $\xi$ in Eq.(\ref{eq:xi2}) is in fact the properly normalized ratio between the ``noise" $\Delta S_{\perp,{\rm min}}$ and the ``signal"  
$|\langle {\bf S} \rangle|$. In experiments $\Delta S_{\perp,{\rm min}}$ is directly measured by measuring $S_z$ 
after an appropriate state rotation and $|\langle {\bf S} \rangle|$ is separately deduced from the Ramsey fringes contrast.

\subsection{State of the art}
On one hand the most precise atomic clocks using microwave transitions in cold alkali atoms have already reached the 
quantum partition noise limit with atom numbers up to $N=6\times10^5$  \cite{Salomon:1999}. 
On the other hand, very recently a significant amount of spin squeezing, up to $-8$ dB ($\xi^2=10^{-0.8}$) 
\cite{Oberthaler:2010} was measured in dedicated, proof-of-principle 
experiments. In \cite{Vuletic:2010} squeezing was created in a large sample 
of $N=5\times10^4$ atoms with a feedback mechanism in a resonant optical cavity, while in \cite{Oberthaler:2010}
and in \cite{Treutlein:2010} the squeezing was created in smaller samples, of order $N=10^3$, using atomic interactions  
in bimodal condensates.
The ultimate limits of the different paths to spin squeezing are still an open question. Here we concentrate on a dynamical scheme
using interactions in bimodal condensates \cite{Oberthaler:2010,Treutlein:2010,Sorensen:2001,Poulsen:2001}
and analyze in particular the influence of dephasing, decoherence and non-zero temperature on this squeezing scheme. 

\subsection{Two-mode scalings without decoherence}
\begin{figure}
\centerline{\includegraphics[width=4cm,clip=]{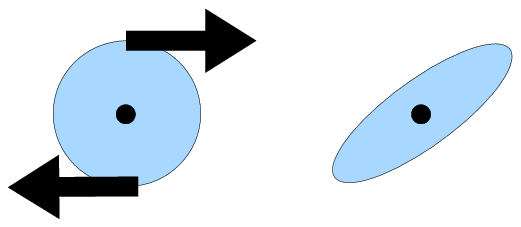}\includegraphics[width=4cm,clip=]{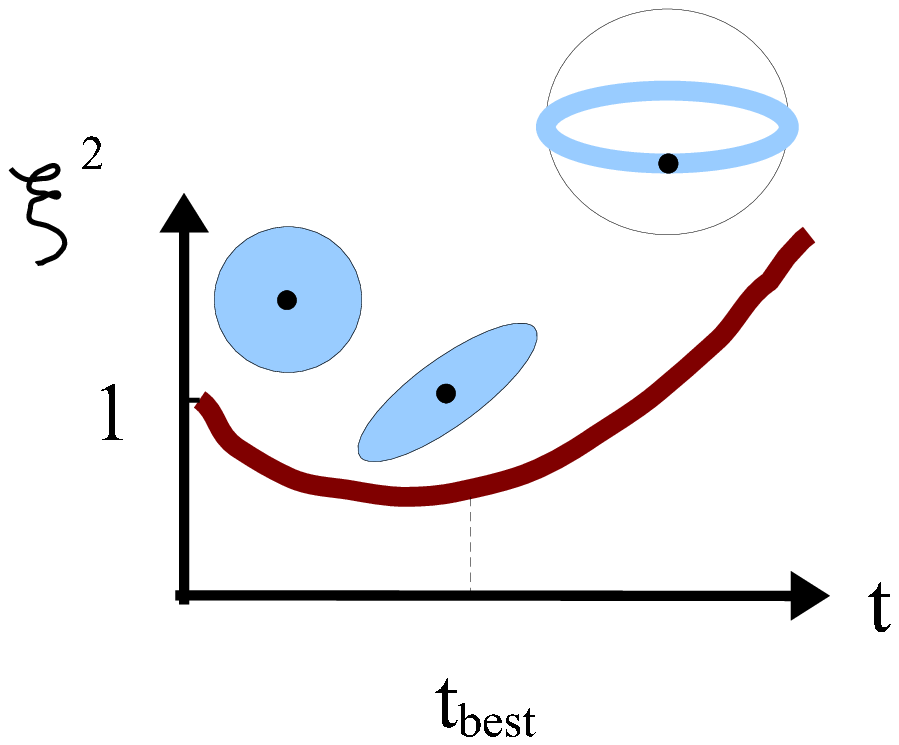}}
\caption{(Color online) Left: Deformation of quantum fluctuations due to the non-linear Hamiltonian (\ref{eq:Hnl}). Right: Best squeezing time.
\label{fig:distorted}}
\end{figure}

We consider for simplicity a bimodal condensate with identical interactions in the components $a$ and $b$ with coupling constants 
$g_{aa}=g_{bb}=g$
and no crossed $a$-$b$ interactions \footnote{In ${}^{87}$Rb atoms this may be done by spatial separation of the spin states \cite{Treutlein:2010}
or by Feshbach tuning of the $a$-$b$ scattering length \cite{Oberthaler:2010}.}. 
We assume that the initial state is the factorized state (\ref{eq:phase_state}) with $\phi=0$ and a fixed total number of atoms $N$.
In a two-mode picture, interactions introduce a Hamiltonian that is non-linear in the spin operator:
\begin{equation}
H_{\rm nl} = \hbar \chi S_z^2  \label{eq:Hnl}\,.
\end{equation}
The quadratic form (\ref{eq:Hnl}) is obtained expanding the system Hamiltonian to second order around 
 the average numbers of particles in components $a$ and $b$, $\bar{N}_a$ and $\bar{N}_b$, both equal to $N/2$ for the initial state (\ref{eq:phase_state}) \cite{Castin:1997,Sinatra:1998}.
$\hbar \chi$ is thus the derivative of the chemical potential with respect to the particle number in each component 
$\hbar \chi=d \mu_a/d N_a=d \mu_b/d N_b$ evaluated in $N_a=\bar{N}_a$, $N_b=\bar{N}_b$.
The general expression of the expanded Hamiltonian including drift terms for non-symmetric interactions, non-symmetric splitting or
fluctuations in the total particle number can be found in \cite{Sinatra:1998,YunLi:2009}. 
We are interested in the best squeezing that can be obtained in the thermodynamic limit for a spatially homogeneous system:
\be
N \to \infty\,, \hspace{0.5cm} \rho={N \over V} = {\rm constant},
\ee
we therefore explicitly write $\chi$ in terms of the interaction constant $g$ and the volume $V$ of the system:
\be
\mu_a=\frac{g N_a}{V} \,, \hspace{0.5cm} \mu_b=\frac{g N_b}{V}\,,  \hspace{0.5cm} \chi=\frac{g}{\hbar V} \,. \label{eq:mua_mub_2m}
\ee
We can consider the non-linear Hamiltonian (\ref{eq:Hnl}) as a Hamiltonian of the form (\ref{eq:H0}) with a Larmor frequency $\omega_{ab}$ 
that depends itself on $S_z$. As explained in \cite{Ueda:1993} and shown in Fig.\ref{fig:distorted} (left), $H_{\rm nl}$ ``twists" the transverse
spin fluctuations and generates spin squeezing.
However, in order to minimize $\xi^2$, the evolution should not go to far. E.g.\ when the fluctuations become too much distorted 
and start to wrap around
the Bloch sphere, the ``signal" $|\langle {\bf S} \rangle|$ in the denominator of $\xi^2$ 
decreases and the squeezing parameter increases again.
A sketch of the time dependence of $\xi^2$ when the state (\ref{eq:phase_state}) evolves under the influence of $H_{\rm nl}$ (\ref{eq:Hnl}) is given in
Fig.\ref{fig:distorted} (right). We name ``best squeezing time" $t_{\rm min}$ the time that minimizes $\xi^2(t)$ and 
``best squeezing" $\xi^2_{\rm min}$ the corresponding squeezing.
The analytical expression for the squeezing as a function of time is given in \cite{Ueda:1993}. 
Introducing an appropriate rescaling of the time variable as in \cite{LiYun:2008}, this gives the following scalings of the best squeezing and the best squeezing time for $N\gg1$ and 
$\rho$,$g$ constants:
\be
\hspace{0.3cm} (\xi^2_{\rm min})_{\rm H_{\rm nl}} \simeq {3^{2/3}\over 2}{1\over N^{2/3}}\;;\hspace{0.5cm} 
\left({\rho g  t_{\rm min}\over \hbar }\right)_{\rm H_{\rm nl}} \simeq  3^{1/6} N^{1/3} \,. \label{eq:twomodescalings}
\ee   
This differs from the original prediction in \cite{Ueda:1993} by numerical factors.
Subsequent studies \cite{Sorensen:2001,Sorensen:2002} gave indications 
that the achievement of large squeezing in condensates should be possible even in presence of decoherence, but were not able to confirm or disprove the scalings (\ref{eq:twomodescalings}).

\subsection{New scalings in presence of decoherence}
We will show in sections \ref{sec:losses} and \ref{sec:temperature} that the scalings (\ref{eq:twomodescalings}) 
are disproved when decoherence coming from particle losses or non-zero 
temperature is included in the description. To summarize, instead of tending to zero when $N\to \infty$ in the thermodynamic limit,
$\xi^2_{\rm min}$ tends to a positive constant 
\be
\xi^2_{\rm min} \stackrel{\rm lim. therm.}{\to}{\rm constant}>0\;\;\;.
\label{eq:true_scaling_xi2}
\ee
Concerning the best squeezing time $t_{\rm min}$ we distinguish two cases. In the case of particle losses
$t_{\rm min}$ is finite in the thermodynamic limit and scales as
\be 
{\rho g t_{\rm min} \over \hbar } \propto {1 \over \sqrt{\xi_{\rm min}^2}}\,.
\label{eq:true_scaling_t_min_loss}
\ee
In the case of finite temperature $t_{\min}$ cannot be calculated within our analytical treatment that neglects interactions among Bogoliubov modes.
Nevertheless, we introduce a ``close-to-best" squeezing time $t_{\rm \eta}$ defined in equation (\ref{eq:xi_eta}), at which $\xi^2$ approaches
$\xi_{\rm min}^2$ with a finite precision $\eta$, that scales as
\be 
{\rho g t_\eta \over \hbar } \propto {1 \over \sqrt{\eta \xi_{\rm min}^2}}\,
\label{eq:true_scaling_t_eta}
\ee
provided that $t_{\rm \eta}$ remains smaller than the typical collision time among Bogoliubov modes.

We will show that these new scalings in presence of decoherence are quite general and that the physics of how decoherence acts
is caught by a very simple dephasing model that we shall solve exactly and study in detail in the next section. 

\section{Dephasing Model}
\label{sec:dephasing}

In this section we consider a dephasing Hamiltonian model of the form
\be
H=\hbar \chi ( S_z^2 + D S_z) \label{eq:Hdephasing}
\ee
where $D$ is a Gaussian real random variable of zero mean. 
We assume here that $D$ is time independent, but it varies
randomly from one experimental realization to the other mimicking a stationary random dephasing environment. 
We also assume that $D$ has a variance of the order of $N$ for $N$ large 
\footnote{One could ask what would be the scaling of $\langle D^2 \rangle$ in the thermodynamic limit
in case the dephasing would have a real physical origin. Let us consider for example an external random potential $\delta U({\bf r})$
with zero mean, $\langle \delta U({\bf r}) \rangle=0$,
that would induce an opposite energy shift for the two components: within
the two-mode model,  
$\delta H_a= \int d^3r |\phi_a|^2 \delta U({\bf r})  a^\dagger a$ and $\delta H_b= -\int d^3r |\phi_b|^2 \delta U({\bf r})  b^\dagger b$.
In this case, from equations (\ref{eq:Hdephasing}) and (\ref{eq:mua_mub_2m}) one has
$\langle (gD)^2 \rangle = \int d^3r \int d^3r^\prime \langle \delta U({\bf r}) \delta U({\bf r^\prime}) \rangle$.
The scaling (\ref{eq:epsnoise}) would then correspond to a fluctuating potential $\delta U({\bf r})$ that has short-range spatial correlations
so that the correlation function $\langle \delta U({\bf r}) \delta U({\bf r'}) \rangle=f(|{\bf r}-{\bf r'}|)$ is
integrable. On the other hand, a uniform fluctuating potential $\delta U$ would correspond to $\langle D^2 \rangle$ scaling as $N^2$. One might
finally imagine a fluctuating potential $\delta U({\bf r})$ that is non-zero only in a finite region of space. In this case
one would get $\langle D^2 \rangle$ independent of $N$. This is what we call the ``weak dephasing limit". We treat this limit in subsection
\ref{sub:weakdephasing}.}:
\be
{\langle D^2 \rangle \over N} \to {\epsilon_{\rm noise}} \:, \hspace{0.5cm} N \to \infty. \label{eq:epsnoise}
\ee
Finally $\epsilon_{\rm noise}$, finite in the thermodynamic limit, 
is a small parameter of the theory and we limit ourselves in general to first order in this quantity. An exception is made 
in subsection \ref{sub:exact} where expressions to all orders in $\epsilon_{\rm noise}$ are given.

Starting with the initial state (\ref{eq:phase_state}) with $\phi=0$, we will show that this
minimal model reproduces the scalings (\ref{eq:true_scaling_xi2}) and (\ref{eq:true_scaling_t_eta}). 
In the subsequent sections \ref{sec:losses} and \ref{sec:temperature} 
we will detail an analogy between the dephasing model and microscopic models
accounting for the effect of particle losses or of non-zero temperature on squeezing. 
The parameter $\epsilon_{\rm noise}$ introduced here (\ref{eq:epsnoise}) will then be related to the lost fraction of particles or the populations of thermally excited modes, respectively. 

\subsection{Squeezing in the thermodynamic limit}
For the symmetric case we consider,
the mean spin is always aligned along $x$. The minimum transverse spin variance is
\begin{multline}
\Delta S_{\perp,{\rm min}}^2={1\over 2}\Big[ \langle S_y^2 \rangle+\langle S_z^2 \rangle - \\
\sqrt{(\langle S_y^2 \rangle-\langle S_z^2\rangle )^2+\langle \{S_z,S_y\}\rangle^2} \; \Big] \,,
\end{multline}
where the expectation values $\langle \ldots \rangle$ represent the average over the quantum state and over the random variable $D$.
The notation $\{,\}$ stands for the anticommutator.
Introducing quantities $A$ and $B$,
\bea
A&=&\langle S_y^2 \rangle-{N\over 4} \label{eq:A} \\
B&=&\langle \{S_z,S_y\}\rangle \label{eq:B} \\
\Delta S_{\perp,{\rm min}}^2&=& {1\over 2}\left[ {N\over 2} + A - \sqrt{A^2+B^2} \right] \,. \label{eq:DeltaSmin}
\eea
To derive the scalings (\ref{eq:true_scaling_xi2}) and (\ref{eq:true_scaling_t_eta}), and to have a physical insight, it is convenient to reason in terms of the phases
of the operators $a$ and $b$:
\begin{eqnarray}
a&=&e^{i {\theta}_a} \sqrt{{N}_{a}} \:\: \:\: ,  \:\:\:\: [{N}_{a},{\theta}_{a}]=i \\
b&=&e^{i {\theta}_b} \sqrt{{N}_{b}} \:\: \:\: ,  \:\:\:\: [{N}_{b},{\theta}_{b}]=i \,,
\end{eqnarray}
where $N_a=a^\dagger a$ and $N_b=b^\dagger b$.
This is a legitimate representation as long as the condensate modes have a negligible probability of being empty \cite{Carruthers:1968}. 
By neglecting the fluctuations of their modulus, the collective spin components $S_x$, $S_y$ are simply given by
\bea
S_x&\simeq&{\rm Re} \: {N\over 2}e^{-i(\theta_a-\theta_b)}\;, \label{eq:simpleSx} \\
S_y&\simeq&{\rm Im} \: {N\over 2}e^{-i(\theta_a-\theta_b)}\;. \label{eq:simpleSy} 
\eea
At $t=0$, the phase difference $(\theta_a-\theta_b)$ has zero mean and root mean square fluctuations that scale as $1/\sqrt{N}$.
Both $S_y$ and $S_z$ scale as $\sqrt{N}$. As a consequence, for $N$
large we can expand the exponentials in (\ref{eq:simpleSx})-(\ref{eq:simpleSy}). To lowest order we then have
\be
S_x \simeq {N\over 2}\;, \hspace{0.25cm} S_y \simeq - {N\over 2} (\theta_a-\theta_b)\;,
\hspace{0.25cm} S_z = {N_a-N_b\over 2}  \,. \label{eq:simple2SxSySz}
\ee
$S_y$ and $S_z$ are then simply proportional to the position operator $Q$ and the momentum operator $P$ of a fictitious free particle.
As we will see, the expansions (\ref{eq:simple2SxSySz}) remain valid for times $(\rho g t/\hbar) \ll \sqrt{N}$.
The squeezing occurs because in a given realization of the experiment $S_y$ becomes an enlarged copy of $S_z$.
Indeed after the pulse, for $t>0$, from the Heisenberg equations of motions for the phase operators, with $\chi=g/(\hbar V)$, one has 
\be
(\theta_a-\theta_b)(t)=(\theta_a-\theta_b)(0^+) - \frac{g t}{\hbar V} \left[ 2 S_z + D \right] \,. \label{eq:rel_phase_delta}
\ee
As the squeezing dynamics goes on, $S_y$ that was initially of the same order as $S_z$, grows linearly in time while $S_z$ stays constant.
Correspondingly $A \propto t^2$ and $B \propto t$.
Since $\langle \{S_y(0),S_z \} \rangle=0$, one actually has at all times:
\bea
S_y&=&S_y(0)+S_y^{\rm lead} \, t\,;  \\
{A\over N} &=& \alpha t^2 \,;  \\
{B \over N}&=& \beta t   \,,
\eea
where we have introduced the time independent operator and coefficients
\bea
S_y^{\rm lead}&=&\frac{\rho g}{2 \hbar} \left[ 2 S_z + D \right] \label{eq:Sylead}\\
\alpha &=&{\langle (S_y^{\rm lead})^2 \rangle \over N}=\left( \frac{\rho g}{2 \hbar}\right)^2(1+\epsilon_{\rm noise}) \label{eq:a} \\
\beta &=&{\langle \{ S_y^{\rm lead},S_z\} \rangle \over N}= \frac{\rho g}{2 \hbar} \label{eq:b}\,.
\eea
Using the fact that from (\ref{eq:simple2SxSySz}) $S_x \simeq N/2$ for $(\rho g t/\hbar) \ll \sqrt{N}$, and expanding the expression (\ref{eq:DeltaSmin}) for $t\gg \hbar/(\rho g)$, we have in the thermodynamic limit
\be
\xi^2(t)\underset{t \to \infty}{=} 1- {\beta^2\over \alpha} +   {\beta^4 \over 4 \alpha^3}  {1\over t^2}+ O(t^{-4}) \,. \label{eq:asympt_xi_t}
\ee
Using equations (\ref{eq:Sylead})-(\ref{eq:b}), with $\epsilon_{\rm noise}\ll 1$, we finally obtain  in the long time limit
\be
\xi^2(t)={\epsilon_{\rm noise}}  + \left({\hbar\over \rho gt}\right)^2\left[ 1 + O\left( {\epsilon_{\rm noise}} \right)\right] +
O\left(  {\hbar^4 \over (\rho gt)^4} \right). \label{eq:xi2t_decoh}
\ee
\subsubsection{Best squeezing and close-to-best time}
According to (\ref{eq:xi2t_decoh}), the best squeezing in the thermodynamic limit to leading order in $\epsilon_{\rm noise}$ is 
\be
\xi^2_{\rm min}=\epsilon_{\rm noise}={\rm lim}_{N\to \infty } {\langle {D}^2 \rangle \over N} \label{eq:xi2Delta} \,.
\ee
Remarkably, the best squeezing (\ref{eq:xi2Delta}) only involves the part of the phase difference $D$ that 
is not proportional to $S_z$. 
To understand physically this result we rewrite
\be
\xi^2_{\rm min}= 1-  {\beta^2\over \alpha}=\frac{\langle (S_y^{\rm lead})^2\rangle \langle S_z^2 \rangle -
 \langle \{S_y^{\rm lead}, S_z\}/2 \rangle^2}
{\langle (S_y^{\rm lead})^2 \rangle \langle S_z^2 \rangle} \,. \label{eq:xi2S}
\ee
Due to interactions, through  the phase difference (\ref{eq:rel_phase_delta}), 
$S_y^{\rm lead}$ is proportional to $(2S_z+D)$.  In the absence of the term $D$ 
this allows a perfect cancellation between the correlation $\langle S_y^{\rm lead} S_z \rangle^2$ and the
product $\langle (S_y^{\rm lead})^2 \rangle \langle S_z^2 \rangle$ in (\ref{eq:xi2S}) leading to $\xi_{\rm min}^2=0$ in the limit
$N\to \infty$. In presence of ${D}$, this is not possible and $\xi^2_{\rm min}$ has a non-zero limit.

Considering the next to leading order in the time expansion of the squeezing parameter Eq.(\ref{eq:xi2t_decoh}),
the best squeezing is reached in an infinite time (in the thermodynamic limit).
However as we will see $\xi^2(t)$ is quite flat around its minimum, and it
  suffices to determine a ``close-to-best" squeezing time $t_\eta$ defined as
\be
 \xi^2(t_\eta)=(1+\eta) \xi_{\rm min}^2\:, \hspace{0.5cm} \eta>0 \,. \label{eq:xi_eta}
 \ee 
  Then, according to (\ref{eq:xi2t_decoh}),
  $t_\eta$ is given by
\begin{equation}
\frac{\rho g}{\hbar} t_\eta = {1 \over \sqrt{\eta \xi^2_{\rm min}}} \,.\label{eq:t_best_xi2}
\end{equation}
The close-to-best squeezing time $t_\eta$ is thus very simply related to the best squeezing $\xi^2_{\rm min}$.
The important point is that $t_\eta$ is finite (non-infinite and non-zero) in the thermodynamic limit.

\subsubsection{Geometrical interpretation}
\begin{figure}
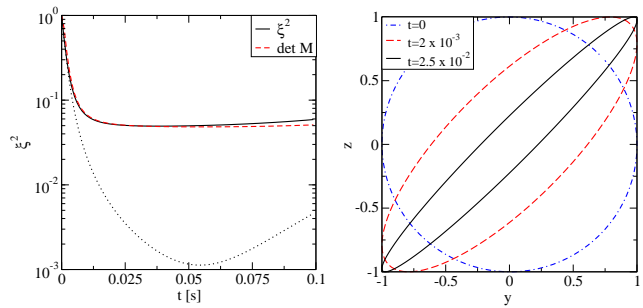

\centerline{\includegraphics[width=4.2cm]{fig_t.eps}\hspace{0.2cm}\includegraphics[width=4cm]{ellipses.eps}}
\caption{(Color online) Left: Squeezing parameter $\xi^2$ as a function of time for $\epsilon_{\rm noise}=0.05$ (full line)
and $\epsilon_{\rm noise}=0$ (dotted line). Determinant of the covariance matrix $M$ as a function of time (dashed line). 
Simulation with $10^4$ realizations. $N=4\times10^4$, $\chi=0.022577$s${}^{-1}$. Right: Isocontours of $W(y,z)$ for 
$t=0$, $t=2\times10^{-3}$s and $t=2.5\times10^{-2}$s.
\label{fig:ellipse}}
\end{figure}

We give here a geometrical and pictorial interpretation to the squeezing process in the thermodynamic limit.
Let us introduce the rescaled transverse spin components
\be
Y={S_y\over \sqrt{\langle S_y^2 \rangle}} \,, \hspace{0.5cm} Z={S_z \over \sqrt{\langle S_z^2 \rangle}} 
\ee
that verify $\langle Y \rangle=\langle Z \rangle=0$ and $\langle Y^2 \rangle=\langle Z^2 \rangle=1$.
At short times, the Wigner function representing the probability distribution of $Y$ and $Z$ is approximately  Gaussian
\be
W(y,z) \propto {\rm exp} \left[ - {1\over 2} \, (y,z) \, M^{-1} \left(\begin{array}{c}y \\ z \end{array}\right) \right]
\ee
where $M$ is the covariance matrix
\be
M= \left(\begin{array}{cc}  \langle Y^2 \rangle & {1  \over 2} \langle \{Y,Z\} \rangle \\  
{1  \over 2} \langle \{Y,Z\} \rangle &   \langle Z^2 \rangle \end{array}\right) \,.
\ee
We can represent graphically the fluctuations of $Y$ and $Z$ by drawing isocontours of $W(y,z)$. 
Let us introduce the eigenvalues of $M$
\be
\lambda_{1,2}=1 \pm {1  \over 2} \langle \{Y,Z\} \rangle \,.
\ee
and a rotated coordinate system $y'-z'$ aligned with the eigenvectors of $M$:
\be
y'=(z+y)/\sqrt{2} \;\;; \hspace{0.5cm} z'=(z-y)/\sqrt{2} \,.
\ee
The points in the $y'z'$ plane such that 
\be
{y'^2 \over \lambda_1} + {z'^2 \over \lambda_2} =1 \label{eq:iso}
\ee
form an ellipse whose semi-axis gives the mean square fluctuations of $Y'$ and $Z'$ that are time dependent linear combinations
of $S_{y}$ and $S_{z}$.
The ellipse surface ${\cal S}_{\rm ellipse}=\pi (\lambda_1 \lambda_2)^{1/2}$ divided by $\pi$ is equal to the square root of the determinant of $M$
and is thus asymptotically equivalent to $\xi_{\rm min}$ according to  (\ref{eq:xi2S}):
\be
\xi_{\rm min}={{\cal S}_{\rm ellipse} \over \pi}  \hspace{0.25cm}\mbox{for}\hspace{0.25cm} t \gg {\hbar \over \rho g} \,.
\ee
In Fig.\ref{fig:ellipse} (left) we show the time dependence of the squeezing parameter $\xi^2$ in presence and in absence of decoherence.
On the same plot we show the determinant of $M$ that asymptotically gives the value of $\xi^2_{\rm min}$  (\ref{eq:xi2S}).
In Fig.\ref{fig:ellipse} (right) we show the isocontours of $W(y,z)$ defined by (\ref{eq:iso}) at different times for the case with decoherence. 
As the dynamics goes on, $Y$ and $Z$ become more and more correlated and the ellipse shrinks.
In the absence of dephasing and in the thermodynamic limit the ellipse would collapse into a segment in the $y=z$ direction. 
In the presence of decoherence the process is ``blocked" and the ellipse keeps a finite width with a limit area 
${\cal S}_{\rm ellipse}=\pi \epsilon_{\rm noise}$. 

\subsection{Exact solution of the dephasing model}
\label{sub:exact}

The dephasing model (\ref{eq:Hdephasing}) is exactly solvable. One first writes the Heisenberg equations of motion for $a$ and $b$, e.g.
\be
i \, \dot{a} =  {\chi \over 2} \left(2 S_z +D +{1 \over 2}\right) a \,.
\ee
Then one uses the fact that $S_z$ is  a constant of motion to integrate the equations \cite{Wright:1997,Sinatra:1998}. One obtains (see also \cite{Minguzzi:arxiv}):
\be
\xi^2(t)={{N\over 2} \left[{N\over 2}+A-\sqrt{A^2+B^2}\right] \over C^2 } \label{eq:xi2_JC}
\ee
with
\bea
A&=&{N(N-1)\over 8} \left[1-e^{-2 (\chi t)^2 \langle D^2 \rangle} (\cos 2 \chi t)^{N-2} \right] \\ \label{eq:A_JC}
B&=&{N(N-1)\over 2} \sin \chi t \; e^{-{1\over 2}(\chi t)^2 \langle D^2 \rangle } \left(\cos  \chi t\right)^{N-2} \\ \label{eq:B_JC}
C&=& \langle S_x \rangle= {N\over 2} e^{-{1\over 2}(\chi t)^2 \langle D^2 \rangle } \left(\cos \chi t\right)^{N-1} \,. \label{eq:C_JC}
\eea
A first application of the exact solution (\ref{eq:xi2_JC})-(\ref{eq:C_JC}) 
is to determine the best squeezing  $\xi_{\rm min}^2$ in the thermodynamic limit,
to all orders in the dephasing parameter $\epsilon_{\rm noise}$. 
To this aim we take the limit $N\to \infty$ in (\ref{eq:xi2_JC})-(\ref{eq:C_JC}) at fixed time $t$, density $\rho$ and
noise parameter $\epsilon_{\rm noise}={\langle D^2 \rangle} / N$. We find that $A/N$, $B/N$ and $C/N$ have a finite limit and that
\begin{multline}
\xi^2(t) \stackrel{\rm lim. therm.}{\to} 1 - \left[ {1\over 2} \left( 1 +   \epsilon_{\rm noise} \right) + 
\phantom{\sqrt{{1\over 4} \left( 1 +   \epsilon_{\rm noise} \right)^2 + \left({\hbar \over \rho g t}\right)^2}}  \right. \\ \left.
\sqrt{{1\over 4} \left( 1 +   \epsilon_{\rm noise} \right)^2 + \left({\hbar \over \rho g t}\right)^2} \; \right]^{-1}  \,. \label{eq:xi2t_exact}
\end{multline}
From this solution one gets the best squeezing and the close-to-best squeezing time $t_\eta$ (for $\eta \, \epsilon_{\rm noise}<1$):
\bea
\xi_{\rm min}^2 &\stackrel{\rm lim. therm.}{\to}& {\epsilon_{\rm noise} \over 1 + \epsilon_{\rm noise}}  
\label{eq:xi2Delta_exact} \\
{\rho g  \over \hbar} t_\eta &\stackrel{\rm lim. therm.}{\to}& \frac{1-\eta \, \epsilon_{\rm noise} }{ (1 + \epsilon_{\rm noise}) \sqrt{\eta \, \epsilon_{\rm noise} }  } \,.
\label{eq:teta_exact}
\eea
Note that one can obtain (\ref{eq:xi2Delta}) and (\ref{eq:t_best_xi2}) from (\ref{eq:xi2Delta_exact}) and (\ref{eq:teta_exact}) 
by linearizing for small $\epsilon_{\rm noise}$. 

\begin{figure}
\centerline{\includegraphics[width=8cm,clip=]{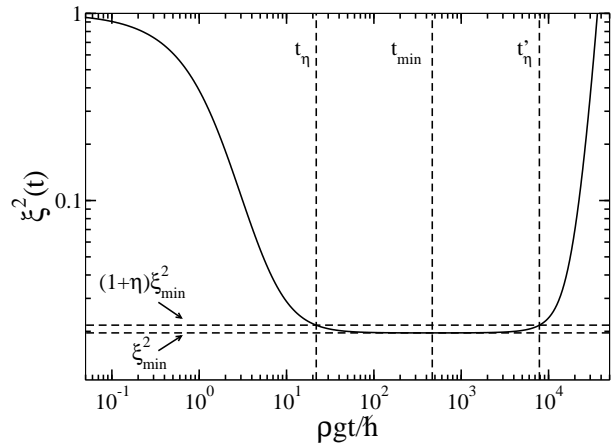}}
\caption{Squeezing as a function of time for the dephasing model (\ref{eq:Hdephasing}) as given
by the exact solution (\ref{eq:xi2_JC}). $N=10^9$, $\langle D^2 \rangle/N=0.02$.
Horizontal dashed lines: $\xi_{\rm min}^2$ given by (\ref{eq:xibest_plus}) and $(1+\eta)\xi_{\rm min}^2$ with $\eta=0.1$.
Vertical dashed lines from left to right: Close-to-best squeezing time $t_\eta$ (\ref{eq:teta_exact}), best squeezing time $t_{\rm min}$ (\ref{eq:tbest}) and  
$t_\eta^\prime$ deduced from (\ref{eq:xi2TP}).
\label{fig:Nunquart}}
\end{figure}

In Fig.\ref{fig:Nunquart} we show $\xi^2(t)$ as a function of time for a large atom number. The curve is indeed quite flat
around the best squeezing time $t_{\rm min}$. There are two solutions to equation (\ref{eq:xi_eta}): $t_{\eta}<t_{\rm min}$
and $t_{\eta}^\prime>t_{\rm min}$. When $N\to \infty$,  $t_\eta$ is finite and given by (\ref{eq:teta_exact}). On the other
hand, as we show in Appendix \ref{app:divergence}, $t_{\rm min}$ diverges as $N^{1/4}$ and $t_{\eta}^\prime$ diverges as $N^{1/2}$.  
Knowing the asymptotic behavior of $t_{\rm min}$, by introducing appropriate rescalings of the time variable as in \cite{LiYun:2008}, it is also possible to
obtain the first finite size correction to $\xi^2_{\rm min}$. This is given in equation (\ref{eq:xibest_plus}) of Appendix \ref{app:divergence}.

\subsection{Squeezing in the weak dephasing limit}
\label{sub:weakdephasing}
We can use the exact solution of the dephasing model (\ref{eq:xi2_JC})-(\ref{eq:C_JC}) to investigate the weak dephasing limit
that we define as the limit where the noise $D$ remains bounded for $N\to \infty$ (see the footnote before equation (\ref{eq:epsnoise})):
\be
{\langle D^2 \rangle } \to {\rm constant} \hspace{0.5cm} N \to \infty.
\ee
In this case the scaling of the best squeezing time with $N$ is still given by $\rho g t_{\rm min}/\hbar \propto N^{1/3}$ 
as in the case without decoherence (\ref{eq:twomodescalings}) and, using the same rescaling of time as in \cite{LiYun:2008}, one has
\bea
\xi^2_{\rm min} &=&{3^{2/3}\over 2}{1\over N^{2/3}} +\frac{\frac{3}{2} + \langle D^2 \rangle}{N} +o\left({1\over N}\right) \,, \\
{\rho g t_{\rm min}\over \hbar} &=& 3^{1/6} N^{1/3} - {\sqrt{3} \over 4} +o(1) \,.
\eea

\section{Particle losses}
\label{sec:losses}
In this section we consider particle losses that are an intrinsic source of decoherence in condensed gases. 
Among those, one-body losses are due to collisions of condensate atoms with residual hot atoms 
due to imperfect vacuum. More fundamental in dense samples are three-body losses where, after a three-body collision, two atoms form a 
molecule and the third atom takes away the energy to fulfill energy and momentum conservation. After such a collision event the three
atoms are lost. Three-body losses are present due to the metastable nature of ultra cold gases, whose real ground state at such low temperatures
would be a solid and whose gaseous phase is maintained because the sample is very dilute. Finally two-body losses
can also be  present, caused by two-body collisions that change the internal state of the atoms. For a trapped gas,
we have shown theoretically
\cite{LiYun:2008} that the best achievable squeezing within a two-mode model at zero temperature in presence of one, two and three-body losses can in principle be very large 
(squeezing parameter $\xi^2$ of the order of $10^{-4}$) provided that the harmonic trapping potential is optimized and a careful choice of the internal state of the atoms is made. To realize such conditions that minimize losses remains however an experimental challenge.

In this section we recall the main results of \cite{LiYun:2008} concerning the squeezing in presence of particle losses,
and we use these results to show an analogy between the effect of the losses and the effect of the
dephasing Hamiltonian (\ref{eq:Hdephasing}) in the thermodynamic limit.

\subsection{Monte Carlo wave functions}
\begin{figure}[htb]
\centerline{\includegraphics[width=8cm,clip=]{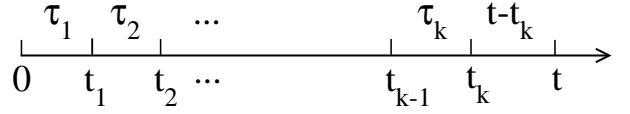}}
\caption{Time sequence of deterministic evolution periods and quantum jumps for a single
Monte Carlo wave function $|\psi\rangle$.}
\label{fig:MCWF_scheme}
\end{figure}

We consider two spatially separated, symmetric condensates. For a more general treatment, please refer to \cite{YunLi:2009}.
Initially the system is in the eigenstate of $S_x$ with maximal eigenvalue $N/2$. 
Besides the non-linear Hamiltonian for the two bosonic modes $a$ and $b$ given by
\be
H_{\rm nl}={\hbar \chi}S_z^2 \hspace{0.25cm}{\rm with}\hspace{0.25cm}  \chi=\left( \partial_{N_a}\mu_a \right)_{\bar{N}_a}/\hbar\,,
\ee
we include one, two and three-body losses. Due to the losses, the system is ``open" and we shall describe it with a 
density operator that obeys a Master Equation of the Lindblad form \cite{Sinatra:1998}.
In the interaction picture with respect to $H_{\rm nl}$:
\be
\frac{d\tilde{\rho}}{dt}=\sum_{m=1}^{3} \sum_{\epsilon=a,b}
\gamma^{(m)}\left[{c}^m_\epsilon\tilde{\rho}{c}^{\dag
m}_\epsilon-\frac{1}{2}\{{c}^{\dag m}_\epsilon{c}^m_ \epsilon,
\tilde{\rho}\} \right] \, \label{eq:me}
\ee
where $c_\epsilon=\tilde{a},\tilde{b}$ for $\epsilon=a,b$ and:
\be
\tilde{a}=e^{{i\over \hbar} H_{\rm nl} t}\, a\, e^{-{i\over \hbar} H_{\rm nl} t} \,;\hspace{0.5cm} \tilde{b}=e^{{i\over \hbar} H_{\rm nl} t} \, b\, e^{-{i\over \hbar} H_{\rm nl} t} \,.
\ee
Here the operators $a$ and $b$ are in the Schr\"odinger picture.
The $m$-body loss rates $\gamma^{(m)}$ are defined in terms of the so-called rate constants $K_m$ as
\be
\gamma^{(m)}=\frac{K_m}{m} \int d^3r|\phi({\bf r})|^{2m}
\ee
where $\phi({\bf r})$ is the condensate wave function in mode $a$ or $b$ for the initial atom number (weak loss approximation), so that for example:
\begin{multline}
{d\over dt}\langle {N}_a \rangle=- \left\langle \left[ K_1  + K_2 N_a \int d^3r|\phi(r)|^{4} + \right.\right. \\ \left. \left.
 K_3 N_a^2  \int d^3r|\phi({\bf r})|^{6} \right] N_a  \right\rangle \,.
\end{multline}
It is convenient to rephrase the Master Equation (\ref{eq:me}) in terms of Monte Carlo wave functions \cite{Molmer:1993}.
In this picture pure states evolve deterministically under the influence of an effective Hamiltonian 
$H_{\text{eff}}$ acting during time intervals $\tau_i=t_i-t_{i-1}$ separated by random quantum jumps (described by the jump
operators $S^{}_\epsilon$) occurring at times $t_i$ as illustrated in Fig.\ref{fig:MCWF_scheme}:
\be
H_{\text{eff}} = -  \sum_{\epsilon=a,b} \frac{i\hbar}{2}\gamma^{(m)} c_\epsilon^{\dag m}c_\epsilon^m  \,,
\hspace{0.5cm} S^{}_\epsilon=\sqrt{\gamma^{(m)}}{c}^m_\epsilon \,.
\ee
More precisely the evolution of the non-normalized state vector $|\psi(t)\rangle$ in between quantum jumps is given by
\be
i \hbar {d\over dt} |\psi(t)\rangle = H_{\rm eff}(t) |\psi(t)\rangle 
\ee
and the effect of a quantum jump in the component $\epsilon_i$ at time $t_i$ is
\be
|\psi(t_i^+)\rangle=S_{\epsilon_i}(t_i) |\psi(t_i^-)\rangle \,.
\ee

Quantum averages of any atomic observables $\mathcal{O}$ are obtained by summing over all the possible trajectories of the non-normalized
state vector \cite{LiYun:2008}:
\be
\langle\hat{\mathcal {O}}\rangle=\sum_{k}\int_{0<t_1<t_2<\cdots t_k<t}  \hspace{-1.5cm} dt_1dt_2\cdots dt_k  \:\: \sum_{\{\epsilon_j\}} \:
\langle\psi(t)|\hat{\mathcal {O}}|\psi(t)\rangle \,. \label{eq:aveMC}
\ee

\subsection{Losses randomly kick the relative phase}
\begin{figure}
\centerline{\includegraphics[width=8cm,clip=]{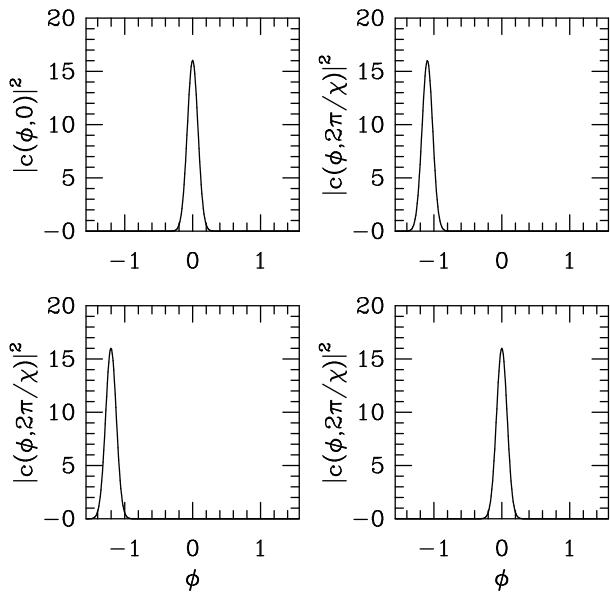}}
\caption{Relative phase probability distribution at time $t=0$ and at time 
$t=2\pi/\chi$ in three single realizations Monte Carlo realization. From upper left to lower right the wave function has
experienced 0, 3, 1 and 0 quantum jumps respectively. Three-body losses corresponding to $N=301$ ${}^{87}$Rb atoms
in $F=1,m_f=-1$ state in separated identical harmonic traps with $\omega=500$Hz. Figure taken from \cite{Sinatra:1998}.
\label{fig:jumps}}
\end{figure}

Let us consider the action of a quantum jump over a phase state (\ref{eq:phase_state})
with $N$ atoms, for example the loss of one particle in state $a$ or $b$ at time $t$:
\bea
c_{a}(t)|\phi \rangle_N &=& \sqrt{{N\over 2}} e^{-i{\chi \over 4}t} e^{i\phi}|\phi - \chi t/2\rangle_{N-1} \,\\
c_{b}(t)|\phi \rangle_N &=& \sqrt{{N\over 2}} e^{-i{\chi \over 4}t} e^{-i\phi}|\phi + \chi t/2\rangle_{N-1} \,.
\eea
Under the action of a jump a phase state remains a phase state. 
On the other hand, the relative phase is shifted  
by a random amount that depends on the time of the jump and has a random sign depending on whether the jump was in $a$ or $b$.
This behavior is illustrated in Fig.\ref{fig:jumps} taken from \cite{Sinatra:1998} where we plot the modulus squared of
a relative phase distribution amplitude $c(\phi,t)$
at $t=0$ and $t=2\pi/\chi$ for three single Monte Carlo realizations. At this particular time (second revival time) the coherent evolution 
due to $H_{\rm nl}$ has no effect and we can isolate the action of the losses.
In the case of Fig.\ref{fig:jumps}, as $\chi t$ is of the order of unity, the shift in the relative phase due a single jump is large.  
In the case of squeezing with $N\gg1$, $\chi t\ll1$, the shifts due to single quantum jumps are on the contrary very small. These
shifts nevertheless limit the maximum squeezing achievable as we shall see.  

\subsection{Spin squeezing limit and the lost fraction}
\begin{figure}[htb]
\centerline{\includegraphics[width=8cm,clip=]{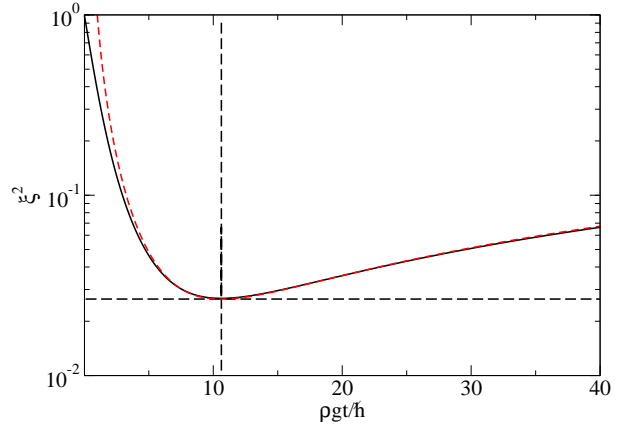}}
\caption{(Color online) Spin squeezing as a function of time in presence of one-body losses:
$\gamma=0.005 \rho g/\hbar$, $N=10^6$. Solid black line: Exact analytical solution. 
Dashed red line: Approximate solution (\ref{eq:xit_losses}). Horizontal and vertical dashed lines: predictions (\ref{eq:sqappr}) 
and (\ref{eq:tappr}) respectively.}
\label{fig:verifxi2} 
\end{figure}

Let us consider the case of one-body losses only with a loss rate constant $\gamma$ equal in states $a$ and $b$. In this case the effective Hamiltonian
does not depend on time;
referring to the time sequence in Fig.\ref{fig:MCWF_scheme}, we have explicitly
\begin{multline}
|\psi(t)\rangle=e^{-iH_{\rm eff}(t-t_k)/\hbar} S_{\epsilon_k}(t_k) e^{-iH_{\rm eff}\tau_k/\hbar} S_{\epsilon_{k-1}}(t_{k-1}) \ldots \\
\ldots S_{\epsilon_1}(t_1) e^{-iH_{\rm eff}\tau_1/\hbar}|\psi(0)\rangle\,,
\end{multline}
and there is an explicit analytical solution for the generated spin squeezing as a function of time
\cite{LiYun:2008,YunLi:2009}. Here we use this solution to find the best squeezing in presence of losses
in the thermodynamic limit with $N\to \infty$, $\rho$ and $\gamma$ constant. For simplicity we also assume
that the fraction of lost particles at the relevant time $t$
remains small in the thermodynamic limit:
\be
\gamma t \equiv \epsilon_{\rm loss} \label{eq:epsilon_loss}\,.
\ee
We proceed similarly as we did to derive Eq.(\ref{eq:xi2t_exact}), taking the thermodynamic limit in the exact solution
in presence of losses.
Restricting for simplicity to the leading order in $\gamma t$ and to the case $\rho g t/\hbar \gg 1$,
we obtain
\bea
{A\over Ne^{-\gamma t}} &\simeq& {\left({\rho g t\over 2\hbar}\right)^2} \left( 1- {5\over 3} \gamma t \right) \\ 
{B\over Ne^{-\gamma t}} &\simeq& {\rho g t \over 2\hbar} \left( 1-  \gamma t \right) \\ 
\xi^2(t) &\simeq& {\gamma t \over 3} + \left({\hbar \over \rho g t} \right)^2 \left[ 1 + O(\gamma t) \right] \,, \label{eq:xit_losses}
\eea
where $A$ and $B$ are defined in (\ref{eq:A})-(\ref{eq:B}).
Equation (\ref{eq:xit_losses}) shows that for long times, the squeezing parameter is asymptotically equivalent 
to one third of the lost fraction of atoms. Minimizing (\ref{eq:xit_losses}) with respect to time we obtain
\bea
\xi_{\rm min}^2 &=& {3\over 4} \left( {4 \over 3} {\hbar\gamma \over \rho g}\right)^{2/3} \label{eq:sqappr} \\
{\rho g \over \hbar} t_{\rm min} &=&  \left( {3 \over 4} {\rho g  \over  \hbar\gamma}\right)^{1/3} \,. \label{eq:tappr}
\eea
In Fig.\ref{fig:verifxi2} we show the squeezing in presence of losses and we compare the exact solution \cite{LiYun:2008}
with the approximate expression (\ref{eq:xit_losses}) valid at long times and in thermodynamic limit. 

\subsection{Analogy between dephasing and losses}
\begin{figure}[htb]
\centerline{\includegraphics[width=8cm,clip=]{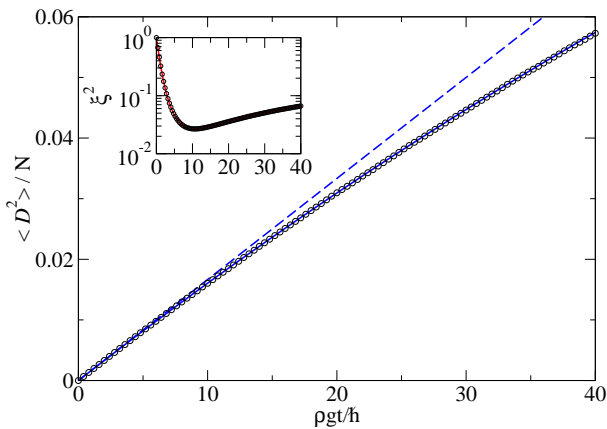}}
\caption{(Color online) Main plot:  $\langle {\cal D}^2 \rangle/N$ given by 
(\ref{eq:calD_pertes}) as a function of time for $\gamma=0.005 \rho g/\hbar$, $N=10^6$. Black circles: Numerical simulation with $5\times 10^5$ 
realizations. 
Dashed blue line: Approximate expression (\ref{eq:approxD2}). Solid blue line: Exact expression
(\ref{eq:exact_d2}). Inset: Squeezing as a function of time.
Solid red line: Exact analytical solution. Black circles: Numerical simulation with $5\times 10^5$ realizations.}
\label{fig:verifD}
\end{figure}

Our goal here is to make an analogy between the dephasing model of section \ref{sec:dephasing} and the present
model with losses. Indeed the relative phase is perturbed by the losses.
In presence of one-body losses, a non normalized Monte Carlo wave function after $k$ jumps has the form:
\bea
|\psi(t)\rangle &=& {\cal N}^{1/2} e^{i\tilde \phi} | \phi + \frac{\chi t}{2} {\cal D} \rangle_{N-k} \,, \\
{\cal N} &=& \left( \prod_{i=1}^k e^{-{\gamma}t_i}\right) e^{-{\gamma}(N-k)t} \gamma^{k} \nonumber \\
&& \times {N(N-1)\ldots(N-k+1) \over 2^k }  \,, \label{eq:norm_pertes} \\
 {\cal D} &=& {1\over t} \sum_{l=1}^k t_l \left( \delta_{\epsilon_l,b}-\delta_{\epsilon_l,a}\right) \,. \label{eq:calD_pertes}
\eea
The factor ${\cal N}$ given by (\ref{eq:norm_pertes}) is the norm squared of the wave function that is needed to calculate quantum averages,
the first line in (\ref{eq:norm_pertes}) is due to the effective Hamiltonian evolution,
$\tilde{\phi}$ is an irrelevant phase and
${\cal D}$ given by (\ref{eq:calD_pertes}) is a random perturbation of the relative phase $2\phi$ that plays the role of the quantity 
$D$ in Eq.(\ref{eq:rel_phase_delta}) of the dephasing model of section \ref{sec:dephasing}, as appears from
the fact that $\exp(-i\chi t D S_z)|\phi\rangle_N = |\phi-\chi t D/2\rangle_N$. Contrarily to $D$, ${\cal D}$ given by 
(\ref{eq:calD_pertes}) is time dependent. 
As detailed in Appendix \ref{app:D2_loss}, using (\ref{eq:aveMC}) we can calculate $\langle {\cal D}^2 \rangle$. To first order 
in $\gamma t$ we obtain:
\be
{\langle {\cal D}^2 \rangle \over N}\simeq {\gamma t \over 3} \, \label{eq:approxD2}\,.
\ee
We have thus shown that $\xi^2(t)$ is asymptotically equivalent to $\langle {\cal D}^2 \rangle/N$ as it the case in the dephasing model.
We can then establish an analogy between the model with losses and the dephasing model as summarized in the first two columns of the Table \ref{tab:correspondence}.
In Fig.\ref{fig:verifD} we show a comparison of  $\langle {\cal D}^2 \rangle/N$ obtained from a Monte Carlo simulation, from the exact
expression (\ref{eq:exact_d2}),  and from the
approximate expression (\ref{eq:approxD2}) valid to first order in $\epsilon_{\rm loss}=\gamma t$.

\subsection{Optimum squeezing in a harmonic trap}
A similar analysis can be performed for a trapped system, where we now consider the two components $a$ and $b$ in identical
and spatially separated harmonic traps. In particular equation (19) of \cite{LiYun:2008} is very similar to (\ref{eq:xit_losses}) except that 
in (19) of \cite{LiYun:2008} we have $N\chi$ instead of $\rho g$ in (\ref{eq:xit_losses}). 
Another difference is that the more general equation (19) in \cite{LiYun:2008} that includes two and three-body losses besides one-body losses, 
is derived performing an approximation on the effective Hamiltonian: the constant loss rate approximation \footnote{
We have verified numerically that this approximation is excellent provided 
that the lost fraction of particles is small.}
\be
H_{\text{eff}} = -  \sum_{\epsilon=a,b} \frac{i\hbar}{2}\gamma^{(m)} c_
\epsilon^{\dag m}c_\epsilon^m  \simeq -  \sum_{\epsilon=a,b} \frac{i\hbar}{2}\gamma^{(m)} {\bar N}^m_\epsilon \,.
\ee

\begin{figure}[htb]
\centerline{\includegraphics[width=8cm]{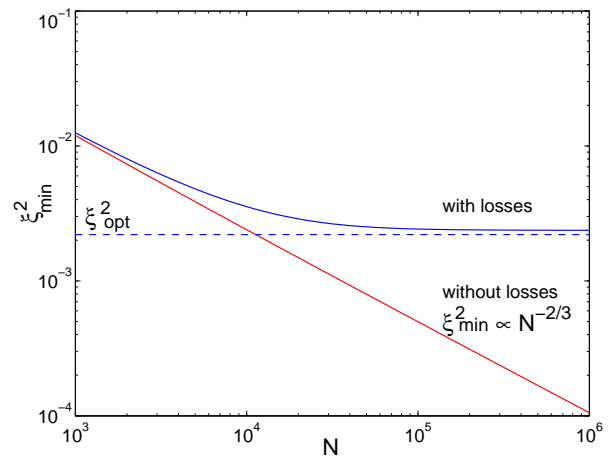}}
\caption{(Color online) Blue line: Best squeezing $\xi_{\rm min}^2$ as a function of $N$ in presence of one,
two and three-body losses, obtained in \cite{LiYun:2008} in the constant loss rate approximation. 
The trap frequency is optimized for each value of $N$ according to (\ref{eq:omega_opt}). 
Dashed blue line: Optimum squeezing for large $N$ given by the analytical prediction (\ref{eq:xibest_trap}).
Red line: Best squeezing without losses for comparison.
Parameters:
$a=5.32$nm, $K^{(1)}=0.1s^{-1}$, $K^{(2)}=2\times 10^{-21}m^3 s^{-1}$, $K^{(3)}=18\times 10^{-42}m^6 s^{-1}$.
This figure is taken from \cite{theseLiYun}.
}
\label{fig:trap_loss}
\end{figure}

By using (19) of \cite{LiYun:2008} and the Thomas-Fermi profiles of the condensate wave functions, 
one can optimize the squeezing with respect to time, trap frequency and atom number.
This optimum squeezing in a trap in presence of one, two and three-body losses
has a simple expression as a function of the $s$-wave scattering length $a_{aa}=a_{bb}=a$ and the rate constants $K_m$,
$m=1,2,3$:
\be
\xi^2_{\rm opt} = \left( \frac{5\sqrt{3}}{28\pi}
\frac{M}{\hbar a} \right)^{2/3}\hspace{-1mm}\left[\sqrt{\frac{7}{2}
(K_1 K_3)}+K_2\right]^{2/3} \label{eq:xibest_trap}\,.
\ee
In Fig.\ref{fig:trap_loss} we show the spin squeezing minimized over time $\xi_{\rm min}^2$ as a function of $N$ in presence of one,
two and three-body losses. The trap frequency is optimized for each value of $N$ \cite{LiYun:2008}:
\be
\omega_{\rm opt} = {2^{19/12} 7^{5/12} \pi^{5/6}\over15^{1/3}}{ \hbar \over m}{a^{1/2}\over N^{1/3}} 
\left( {K_1\over K_3}\right)^{5/12} \,. \label{eq:omega_opt}
\ee
We note that the minimum squeezing, instead of going to zero as in two-mode model without decoherence
(red line) tends to a finite non-zero value for $N\to \infty$
given by (\ref{eq:xibest_trap}). 

\section{Finite temperature}
\label{sec:temperature}

The multimode nature of the atomic field and the population in the excited modes at non-zero temperature
have important consequences on the squeezing and change the scaling laws with respect to the two-mode case (\ref{eq:twomodescalings}).
In \cite{Sinatra:2011} we use a powerful formulation of the Bogoliubov theory in terms of the time dependent condensate 
phase operator \cite{Sinatra:2007,Sinatra:2009} to perform a multimode treatment of the squeezing generation in a condensed gas 
in the homogeneous case. 
We show that the best squeezing $\xi_{\rm min}$ has a finite non-zero value in the thermodynamic limit and we calculate this value analytically.
 
\subsection{Multimode description}
We consider a discretized model on a lattice with unit cell of volume $dV$, within a volume $V$ with periodic boundary conditions 
\cite{Sinatra:2009}. The Hamiltonian after the pulse for component $a$ (and similarly for $b$) reads:
\begin{equation}
H_a= \sum_{\kk} \frac{\hbar^2 k^2}{2m} a^\dagger_\kk a_\kk + 
		\frac{g}{2} dV { \sum_{\rr}}  
	 \psi_a^\dagger(\rr) \psi_a^\dagger(\rr) \psi_a(\rr)  \psi_a(\rr)  \,.
\label{eq:discrHam}
\end{equation}
The fields have commutators
\be
[\psi_\mu(\rr),\psi_\nu^\dagger(\rr')]= {\delta_{\rr \rr'} \delta_{\mu \nu}\over dV}
\ee
with $\mu,\nu=a$ or $b$, and
${a}_\kk ({b}_\kk)$ is the amplitude of ${\psi}_{a,b}$ over the plane wave of momentum $\kk$. 
We assume identical interactions in states $a$ and $b$ with a coupling constant $g=4\pi\hbar^2 a/m$ 
where $a$ is the $s$-wave scattering length while states $a$ and $b$ do not interact $(g_{ab}=0)$.
Note that the coupling constant in $H_{a,b}$ should actually be a {\sl bare} coupling constant $g_0$ different from
the effective coupling constant $g$, but this difference can be made small in the present weakly interacting regime
$(\rho a^3)^{1/2} \ll 1$ by choosing a lattice spacing much larger than $a$ but still much smaller than the healing
length $\propto 1/\sqrt{\rho a}$ \cite{Mora:2003}.

In terms of the fields, the collective spin components are 
\begin{equation}
S_x+i S_y=dV {\textstyle \sum_{\rr}}  \,  {\psi}^\dagger_a
(\rr\,){\psi}_b(\rr\,), \hspace{0.5cm} S_z=\frac{N_a-N_b}{2} \label{eq:Sxyz}
\end{equation}
with $N_\nu=dV \sum_\rr \psi_\nu^\dagger(\rr)\psi_\nu(\rr),$ $\nu=a,b$.

\begin{figure}[tb]
\centerline{\includegraphics[width=8cm,clip=]{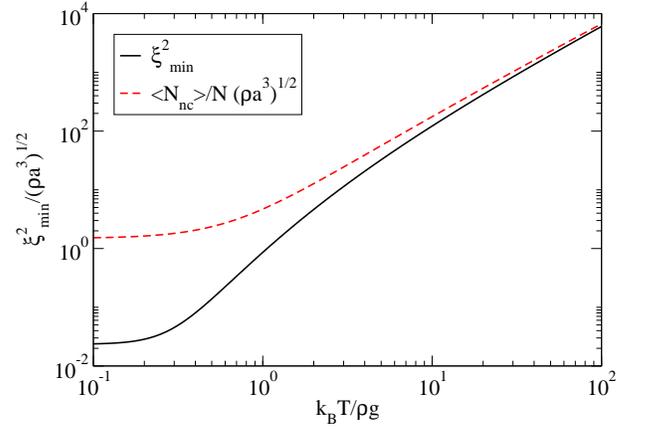}}
\caption{(Color online) $\xi_{\rm min}^2$ (solid line) and before-pulse
non-condensed fraction $\langle N_{\rm nc} \rangle/N$ (dashed line), both divided by $\sqrt{\rho a^3}$, as functions of $k_BT/\rho g$. 
Both quantities are obtained analytically within a Bogoliubov framework, see (\ref{eq:xibest_q}) for $\xi_{\rm min}^2$.}
\label{fig:quant_therm}
\end{figure}
For low or high values of $k_BT/\rho g$ we find the asymptotic behaviors
\bea
{ k_B T \ll \rho g:} &\hspace{0.25cm}& {\xi_{\rm min}^{2}\over \sqrt{\rho a^3}} 
						\simeq{\xi_{\rm min}^{2 \,(T=0)}\over \sqrt{\rho a^3}} = 0.0234\ldots \\
{ k_B T \gg \rho g:} &\hspace{0.25cm}& \xi_{\rm min}^2 \simeq {\langle N_{\rm nc} \rangle \over N} \,.
\eea

\begin{table*}
\label{tab:correspondances}
\begin{tabular}{|c|c|c|}
\hline
{\bf Particle Losses} & {\bf Dephasing model} & {\bf Multimode} $T\neq0$ \\
\hline 
& &\\
$|\psi(t)\rangle = {\cal N}^{1/2} e^{i\tilde \phi} | \phi + \frac{\chi t}{2} {\cal D} \rangle$ 
& $(\theta_a-\theta_b)(t)=(\theta_a-\theta_b)(0^+) - {\chi t} \left[ 2 S_z + D \right]$
& $(\theta_a-\theta_b)(t)=(\theta_a-\theta_b)(0^+) - {\chi t}\left[ 2 S_z + D_{\rm th} \right]$ \\
& &\\
\hline 
& &\\
${\cal D}$ {\small from quantum jumps} (\ref{eq:calD_pertes}) & $D$ {\small from a dephasing Hamiltonian} (\ref{eq:Hdephasing}) &
$D_{\rm th}$  {\small from excited modes population} (\ref{eq:Dth}) \\
 & &\\ 
\hline
& &\\
${ \displaystyle \xi^2(t) \underset{\rho gt/\hbar>1}{\simeq}  {\langle {\cal D}^2 \rangle \over N}}$  & 
   ${ \displaystyle \xi^2(t) \underset{\rho gt/\hbar>1}{\simeq}  {\langle D^2 \rangle \over N}}$  & 
   ${ \displaystyle \xi^2(t) \underset{\rho gt/\hbar>1}{\simeq}  {\langle D_{\rm th}^2 \rangle \over N}}$  \\
    & &\\ 
\hline
 & &\\  
${\displaystyle {\langle {\cal D}^2 \rangle \over N} = {\gamma t \over 3}={\epsilon_{\rm loss} \over 3}}$ & 
  ${\displaystyle {\langle D^2 \rangle \over N} = \epsilon_{\rm noise}}$ &
  ${\displaystyle {\langle D_{\rm th}^2 \rangle \over N} = \sqrt{\rho a^2} F(k_BT/\rho g)}\underset{k_BT > \rho g}{\sim}
  \epsilon_{\rm Bog}$ \\
 & &\\   
\hline  
\end{tabular}
\caption{Correspondence table among the different physical models. The results are valid in the thermodynamic limit and to first order in the 
small parameters $\epsilon_{\rm loss}$ (\ref{eq:epsilon_loss}), $\epsilon_{\rm noise}$ (\ref{eq:epsnoise}) and  
$\epsilon_{\rm Bog}$ (\ref{eq:epsilon_Bog}).}
\label{tab:correspondence}
\end{table*}

\subsection{Best squeezing and close-to-best time} 
Performing a double expansion for large $N$ and small non-condensed fraction $\epsilon_{\rm Bog}$,
\be
\epsilon_{\rm Bog} \equiv {\langle N_{\rm nc} \rangle \over N} \ll 1 \label{eq:epsilon_Bog}\,
\ee
we find that to first order the system effectively behaves as in the dephasing model presented in section \ref{sec:dephasing}.
Indeed the component $S_y$ of the spin develops a term that is proportional to the condensate relative phase $\theta_a-\theta_b$, and 
the relative phase evolves as 
\be
\theta_a-\theta_b= (\theta_a-\theta_b)(0^+) - \frac{gt}{\hbar V} 
\left[ ( N_a - N_b ) + D_{\rm th} \right] \,. \label{eq:temp_theta}
\ee
The dephasing parameter $D_{\rm th}$ in (\ref{eq:temp_theta}) is related to the population in the excited
modes:
\begin{equation}
D_{\rm th} =  \sum_{\kk \neq {\bf 0}} \left( U_k + V_k \right)^2 \left( n_{a \kk} - n_{b \kk} \right) \,,
\label{eq:Dth}
\end{equation}
where $n_{a \kk}=c_{a \kk}^\dagger c_{a \kk}$ and $n_{b \kk}=c_{b \kk}^\dagger c_{b \kk}$ are the occupation number operators
of the Bogoliubov modes. Note that these modes are in a {\sl non-equilibrium} state
since the zero relative phase state  between the condensates in $a$ and $b$ is prepared
at $t=0^+$ by applying a sudden $\pi/2$ pulse to the gas (that was initially at thermal equilibrium
in state $a$). $U_k$ and $V_k$ are the usual Bogoliubov functions 
\be
U_k+V_k=\left( {E_k \over E_k + \rho g}\right)^{1/4} ; \hspace{0.5cm} E_k={\hbar^2k^2 \over 2m} \,, \label{eq:sk}
\ee
$\rho/2=N/2V$ being the spatial density in each single component $a$ or $b$ after the pulse. 
$D_{\rm th}$ fluctuates from one realization to the other because $n_{a \kk}$ and $n_{b \kk}$ depend (in Heisenberg picture)
on the creation and annihilation operators of the Bogoliubov modes before the pulse, and the initial state of the gas has
thermal fluctuations.

To first order in $N$ and in $\epsilon_{\rm Bog}$, the best squeezing parameter
and the close-to-best squeezing time are given by \cite{Sinatra:2011}
\be
\xi_{\rm min}^2 = \frac{\langle D_{\rm th}^2 \rangle}{N} \hspace{0.25cm};\hspace{0.25cm} \frac{\rho g t_\eta}{\hbar} = {1 \over \sqrt{\eta \xi_{\rm min}}}
\,.
\ee
An explicit calculation \cite{Sinatra:2011} gives:
\begin{equation}
\xi_{\rm min}^2 \!=\!\!
\int\!\! {d^3 k \over (2\pi)^3} \, {s_k^4 \over 2\rho} \left[ \Big(n_k^{(0)}+{1\over 2}\Big)\! 
\left( {  (s_k^{(0)})^2\over  s_k^4 } + {s_k^4 \over  (s_k^{(0)})^2 }  \right) -1 \right] \,,
\label{eq:xibest_q}
\end{equation}
where now $n_k^{(0)}$ are Bose mean occupation numbers of Bogoliubov modes in state $a$ before the pulse
\be
n_k^{(0)}={1\over e^{\epsilon_k/k_BT}-1}, \hspace{0.5cm} \epsilon_k=\sqrt{E_k(E_k+2\rho g)} \,.
\ee 
$s_k=U_k+V_k$ and $s_k^{(0)}=U_k^{(0)}+V_k^{(0)}$ where $U_k^{(0)}$ and $V_k^{(0)}$ are the Bogoliubov functions in internal state $a$ before the pulse
\be
U_k^{(0)}+V_k^{(0)}=\left( {E_k \over E_k + 2 \rho g}\right)^{1/4} ; \hspace{0.5cm} E_k={\hbar^2k^2 \over 2m} \,. \label{eq:sk0}
\ee
A consequence of (\ref{eq:xibest_q}) is that the squeezing divided by $\sqrt{\rho a^3}$ is an universal function of $k_BT/\rho g$.

\subsection{Spin squeezing and non-condensed fraction}
Within our treatment, valid for large $N$ and $T\ll T_c$, we find that the best squeezing $\xi^2_{\rm min}$ (\ref{eq:xibest_q}) is always lower than the 
before-pulse non-condensed fraction.
This is shown in Fig.\ref{fig:quant_therm}.

Finally in Table \ref{tab:correspondances} we can complete the correspondence table between dephasing noise, losses and non-zero temperature effects.

\section{A few words about experiments}
Two recent experiments demonstrated spin squeezing in Bose-condensed bimodal condensates of rubidium atoms in internal states
$|a\rangle=|F=1,m_F= \mp1\rangle$ and $|b\rangle=|F=2,m_F=\pm1\rangle$ \cite{Treutlein:2010,Oberthaler:2010}. 
Due to the fact that the three scattering lengths 
characterizing the interactions between atoms are very close
\be
g_{aa} \simeq g_{ab} \simeq g_{bb}
\ee
the effective two-mode nonlinearity $\chi$ in $H_{\rm nl}$ (\ref{eq:Hnl}) is very small when the condensates $a$ and $b$ overlap
spatially. This gives the possibility to {\it tune} the non-linearity \cite{YunLi:2009} by either controlling the spatial overlap between
the two species as done \cite{Treutlein:2010} or by using a Feshbach resonance changing the inter-species coupling constant $g_{ab}$ 
as done in \cite{Oberthaler:2010}. A fundamental source of decoherence in these experiments is two-body losses
in the state $F=2$. A theoretical analysis \cite{YunLi:2009} shows that in typical experimental conditions these losses limit the squeezing to about 
$\xi_{\rm min}^2 \simeq 6  \times 10^{-2}$. For cold samples this limit is above the limit imposed by non-zero temperature.
In the case of \cite{Treutlein:2010} we could explain in detail the squeezing results using a zero-temperature model including spatial dynamics
and including particle losses and technical noise (dephasing noise) as sources of decoherence, the latter being dominant \cite{Treutlein:2010,theseLiYun}.

The perspective of studying experimentally the scaling of squeezing in a controlled decoherence environment, e.g. preparing
the sample at different temperatures, is fascinating and challenging. 

\section{Conclusions}
We have considered a scheme to create spin squeezing using interactions in Bose-condensed gas with two internal states 
\cite{Ueda:1993,Sorensen:2001,Poulsen:2001}. The squeezing is created dynamically after a $\pi/2$ pulse applied on the system initially at equilibrium in
one internal state. We have reviewed the ultimate limits of this squeezing scheme imposed by particle losses and non-zero temperature based on our recent works
\cite{LiYun:2008} and \cite{Sinatra:2011} and we have extracted a simple
physical picture of how decoherence acts in the system. 
An important result is that contrarily to the
case without decoherence \cite{Ueda:1993} the squeezing parameter minimized over time $\xi_{\rm min}^2$ has a finite non-zero value
in the thermodynamic limit, that we determine analytically.
Finally we have shown that the physics of spin squeezing in presence of losses
or at non-zero 
temperature can be caught by a simple dephasing model (also considered in \cite{Minguzzi:arxiv}) that we have solved exactly and studied in details.

\appendix

\section{Times $t_{\rm min}$ and $t_\eta^\prime$ in the dephasing model}
\label{app:divergence}

The exact solution (\ref{eq:xi2_JC})-(\ref{eq:C_JC}) allows to determine how the best squeezing time $t_{\rm min}$ diverges
in the thermodynamic limit. We first found numerically that it diverges as $N^{1/4}$.
We then introduce the rescaled time $\theta$ such that
\be
{\rho g \over \hbar}t = \theta N^{1/4} \,.
\ee
Expanding the functions $\cos^{N-2}(2\chi t)$ and $\exp[-2(\chi t)^2 \langle D^2 \rangle]$ up to terms $O(1/N)$ included in $A$;
linearizing $\sin(\chi t)$ and expanding $\cos^{N-2}(\chi t)$ and $\exp[-(\chi t)^2 \langle D^2 \rangle/2]$ up to terms $O(1/N^{1/2})$ included
in $B$; and expanding $\cos^{N-1}(\chi t)$ and $\exp[-(\chi t)^2 \langle D^2 \rangle/2]$ up to terms $O(1/N^{1/2})$ included in $C$, we obtain
\begin{multline}
\xi^2(t)= {\epsilon_{\rm noise} \over 1 + \epsilon_{\rm noise}} + \\ {1 \over N^{1/2} } 
\left[ \epsilon_{\rm noise} \theta^2 + {1\over \theta^2 (1+ \epsilon_{\rm noise})^3 } \right] +  O\left({1\over N}\right) \,. \label{eq:xi2theta}
\end{multline}
By minimizing (\ref{eq:xi2theta}) over $\theta$ one obtains 
\bea
{\rho g \over \hbar} t_{\rm min} & \stackrel{\rm lim. therm.}{\sim}& \left[ {N \over  \epsilon_{\rm noise} (1+ \epsilon_{\rm noise})^3}\right]^{1/4} \,, \label{eq:tbest} \\
\xi^2_{\rm min}& \stackrel{\rm lim. therm.}{=} &{ \epsilon_{\rm noise} \over 1 +   \epsilon_{\rm noise} }+ \nonumber \\
& &{2 \over N^{1/2} } {\epsilon_{\rm noise}^{1/2} \over \left( 1+\epsilon_{\rm noise} \right)^{3/2}} + O\left({1\over N}\right)
\label{eq:xibest_plus}
\eea

With a similar technique we can determine the divergence of $t'_\eta$ that is  the second solution of equation (\ref{eq:xi_eta})
$t'_\eta>t_{\rm min}$. When $N\to \infty$, $t_{\eta}^\prime$ diverge as $N^{1/2}$. To calculate the prefactor we introduce again a rescaled time
$\rho g t/\hbar = \theta' [N/(1+ \epsilon_{\rm noise})]^{1/2}$ and take the large $N$ limit in (\ref{eq:xi2_JC}) to obtain
\be
\xi^2(t)=e^{\theta'^2}\left[ 1 - {1\over 1+  \epsilon_{\rm noise}} \, {\theta'^2 \over \sinh \theta'^2 }\right] + O\left({1\over N}\right) \,.
\label{eq:xi2TP}
\ee
Solving the transcendental equation $(1+\eta) \xi^2= \xi^2_{\rm min}$ one finds  the large $N$ approximation to $t_{\eta}^\prime$.
For $\eta \ll 1$ and no constraint on the ratio $\eta/\epsilon_{\rm noise}$, we obtain
\be
\theta'^2_\eta \simeq {2 \eta \over 1+ \sqrt{1+{2\eta \over 3 \epsilon_{\rm noise} }}} \label{eq:t_etaprime} \,.
\ee

\section{Calculation of $\langle {\mathcal {D}}^2 \rangle$ in the lossy model}
\label{app:D2_loss}
From the definition of ${\cal D}$ (\ref{eq:calD_pertes}) and the expression of a quantum average
(\ref{eq:aveMC}) in the Monte Carlo wavefunction method,
using the expression of the norm squared of $|\psi(t)\rangle$ of (\ref{eq:norm_pertes}), we obtain
\begin{multline}
\langle \mathcal{D}^2\rangle (t) = e^{-\gamma Nt} \sum_{k=1}^{N} \binom{N}{k}
\gamma^k e^{k \gamma t}
\int_0^{t}dt_1 \ldots \int_0^t dt_k 
\\
\frac{1}{2^k} 
\sum_{\eta_1,\ldots,\eta_k=\pm 1} 
\left(\frac{1}{t} \sum_{i=1}^{k} \eta_i t_i\right)^2
\left(\prod_{i=1}^{k} e^{-\gamma t_i}\right).
\end{multline}
We have introduced the random variables $\eta_i=+1$ for $\epsilon_i=b$ and $\eta_i=-1$ for $\epsilon_i=a$, and we have used the fact
that the integrand is a symmetric function of the jump times $t_i$ to extend time integration from
the ordered domain $0<t_1< \ldots < t_k$ to the hypercube $[0,t]^k$ (also dividing by $k!$). 
The notation $\binom{N}{k}$ represents the usual binomial coefficient
$N!/[k!(N-k)!]$.
We first sum over the variables $\eta_i$:
\be
\frac{1}{2^k} \sum_{\eta_1,\ldots,\eta_k=\pm 1} \left( \sum_{i=1}^{k} \eta_i t_i\right)^2 = \sum_{i=1}^{k} t_i^2
\ee
then we perform the temporal integration to obtain
\bea
\langle \mathcal{D}^2\rangle &=& e^{-\gamma N t } \sum_{k=1}^N \binom{N}{k} k u^k {I_2 \over I_0} \,, \\
I_0&=& \int_0^t \, dt_1 \, e^{-\gamma t_1} = {1-e^{-\gamma t} \over \gamma}\,, \\
I_2&=& \int_0^t \, dt_1 \, t_1^2 e^{-\gamma t_1} \nonumber \\
&=& {1\over \gamma^3} \left\{ 2-e^{-\gamma t}\left[2+2\gamma t + (\gamma t)^2\right]\right\} \,, \\
u&=& \gamma e^{\gamma t} I_0=  e^{\gamma t} -1 \,.
\eea
Taking the derivative with respect to $u$ of the binomial identity
$\sum_{i=0}^{k} \binom{N}{k} u^k = (1+u)^N$, we get the final expression
\be
{\langle \mathcal{D}^2 \rangle\over N} =  \frac{\gamma I_2}{t^2}.
\label{eq:exact_d2}
\ee
Expanding $I_2$ for small $\gamma t$ gives as expected
\be
{\langle {\cal D}^2 \rangle \over N} \simeq  {1\over 3} \gamma t \,.
\ee

\bibliography{paper}
\bibliographystyle{unsrt}

\end{document}